\documentclass[english,amsmath,notitlepage,amssymb,floatfix,hyperscriptaddress,aps,prl,balancelastpage,twocolumn,reprint]{revtex4-2}

\usepackage[utf8]{inputenc}
\usepackage{CJK}
\CJKfamily{gbsn}
\usepackage{graphicx,bm}
\usepackage{amsmath,amssymb,amsthm}
\usepackage{appendix}
\usepackage{array,booktabs}
\usepackage{pgfplots}
\usepackage{microtype}  

\newcommand\non{\nonumber \\}

\newcommand{\br}{{\bf r}}

\newcommand{\rr}{\bm{\mathrm{r}}}

\newcommand{\uu}{\bm{\mathrm{u}}}

\newcommand{\J}{\bm{\mathrm{J}}}
\newcommand{\F}{\bm{\mathrm{F}}}

\newcommand{\gLG}{\gamma_{\rm LG}}

\theoremstyle{definition}

\theoremstyle{remark}

\usepackage{tikz}
\usepackage{pgfplots}
\usepackage{color}
\usepackage[hidelinks]{hyperref}
\hypersetup{
    colorlinks,
    linkcolor={red!50!black},
    citecolor={blue!50!black},
    urlcolor={blue!80!black}
}
\usepackage[bf,FIGTOPCAP,nooneline]{subfigure}

\begin{document}

\begin{CJK*}{UTF8}{gbsn}
\title{Spontaneous currents determine capillary rise in active matter}

\author{Xinyi Dong (董心怡)}
\affiliation{Center for Soft Condensed Matter Physics and Interdisciplinary Research \& School of Physical Science and Technology, Soochow University, 215006 Suzhou, China}

\author{Yongfeng Zhao (赵永峰)}
\email{yfzhao2021@suda.edu.cn}
\affiliation{Center for Soft Condensed Matter Physics and Interdisciplinary Research \& School of Physical Science and Technology, Soochow University, 215006 Suzhou, China}

\date{\today}

\begin{abstract}
The capillary rise of simple passive fluids in a tube is given by Jurin's law of capillary action, which balances surface tension and gravity. For fluids composed of active particles interacting via pairwise forces, a capillary rise was reported despite a negative mechanical surface tension, a phenomenon which remains unexplained. We establish the active form of Jurin's law from the microscopic dynamics. It includes a drag emerging from particle currents that we find responsible for capillary rise. These active currents, alongside negative surface tension, lead to complex and counterintuitive capillary action phenomena that are impossible in equilibrium. In particular the capillary rise of active fluids depends on the shape of the tube, not solely on the tube diameter. 
\end{abstract}

\maketitle 

\end{CJK*}

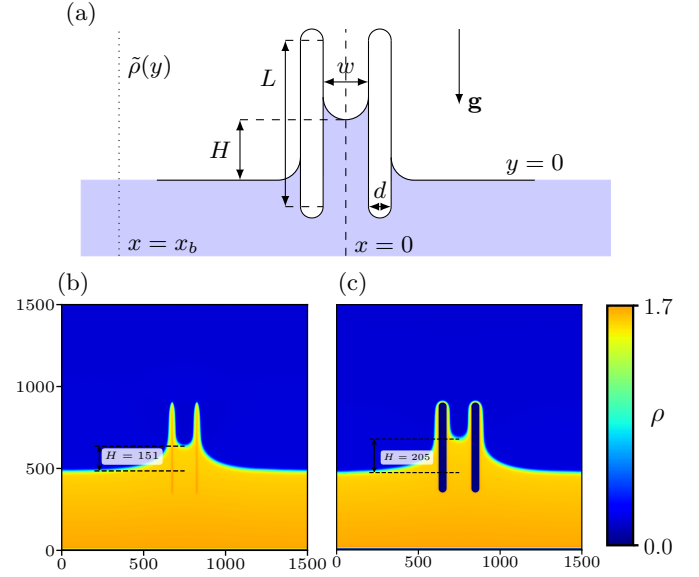
\begin{figure}[t!]
    \centering
    \begin{tikzpicture}[>=latex]
    \begin{scope}[xshift=-2.3cm,yshift=-7.2cm]
        \node at (-3.5,2.7) {(a)};

        \filldraw[blue!20] (-3.5,0.5) rectangle (3.5,-0.5);
        \filldraw[blue!20] (-0.3,1.6) arc(180:360:0.3) -- (0.3,0.5) -- (-0.3,0.5);
        \filldraw[white] (-0.6,0.15) arc(180:360:0.15) -- (-0.3,0.55) -- (-0.6,0.55);
        \filldraw[white] (0.3,0.15) arc(180:360:0.15) -- (0.6,0.55) -- (0.3,0.55);
        \filldraw[blue!20] (-0.9,0.5) arc(270:360:0.3) -- (-0.6,0.5);
        \filldraw[blue!20] (0.6,0.8) arc(180:270:0.3) -- (0.6,0.5);

        \draw[thin] (-0.3,1.6) arc(180:360:0.3); 
        \draw[thin] (-0.3,2.35) arc(0:180:0.15); 
        \draw[thin] (-0.6,0.15) arc(180:360:0.15); 
        \draw[thin] (-2.5,0.5) -- (-0.9,0.5) arc(270:360:0.3);
        \draw[thin] (0.6,2.35) arc(0:180:0.15); 
        \draw[thin] (0.3,0.15) arc(180:360:0.15); 
        \draw[thin] (0.6,0.8) arc(180:270:0.3) -- (2.5,0.5);
        \draw[<->] (-1.4,0.5) -- node[left] {$H$} (-1.4,1.3);
        \draw[<->] (-0.8,0.15) -- node[left,yshift=0.6cm] {$L$} (-0.8,2.35);
        \draw[dashed] (-0.8,0.15) -- (-0.3,0.15);
        \draw[dashed] (-0.8,2.35) -- (-0.3,2.35);
        \draw[<->] (-0.3,1.8) -- node[above] {$w$} (0.3,1.8);
        \draw[<->] (0.3,0.15) -- node[above] {$d$} (0.6,0.15);
        \draw[dashed] (-1.4,1.3) -- (0,1.3);
        \draw[dashed] (0,-0.5) -- (0,2.5);
        \draw (-0.3,2.35) -- (-0.3,0.15);
        \draw (-0.6,2.35) -- (-0.6,0.15);
        \draw (0.3,2.35) -- (0.3,0.15);
        \draw (0.6,2.35) -- (0.6,0.15);

        \draw[dotted] (-3,-0.5) node[right,yshift=0.12cm] {$x=x_b$} -- (-3,2.5);
        \node[anchor=west, black] at (-3, 2) {$\tilde{\rho}(y)$};
        
        \draw[->] (1.5,2.5) -- (1.5,1.5) node[right] {$\mathbf{g}$};
        \node[anchor=west, black] at (2, 0.7) {$y=0$};
        \node[anchor=west, black] at (0, -0.35) {$x=0$};
    \end{scope}
    \end{tikzpicture}
    \begin{tikzpicture}
    \node at (-4.5,2.2) {(b)};
    \node at (-0.8,2.2) {(c)};
    \node at (-3.2,0.2) {\includegraphics[width=0.23\textwidth]{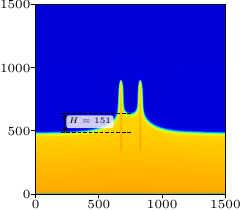}};
    \node at (0.7,0.17) {\includegraphics[width=0.198\textwidth]{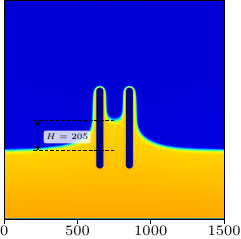}};
    \node at (3,0.3)
        {\includegraphics[width=0.05\textwidth]{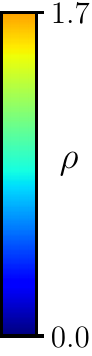}};
      \end{tikzpicture}
    \caption{\textbf{(a)} Setup of capillary action. The system is confined in $y$ and periodic in $x$, the top wall is soft and homogeneous so that liquid droplets do not condensate. Under gravity $\mathbf{g}$, the liquid sits at the bottom of the system. The tube is of width $w$ and length $L$ with impenetrable walls. The tube walls have a hemispherical tip with diameter $d$. We denote $x=0$ at the central axis of tube, and $y=0$ at the liquid level far from the tube. The capillary rise $H$ is defined as the height of the meniscus inside the tube. \textbf{(b-c)} Contrary to an equilibrium system, the capillary rise of an active fluid depend on the shape of the tube, as can be seen by comparing tubes with sharp walls (b) and rounded wall (c).}
    \label{fig:capillary_action}
\end{figure}

Capillary action refers to phenomena where liquids climb up or down a surface against gravity~\cite{de2013capillarity}. In equilibrium, capillary action is controlled by the balance between the liquid-gas surface tension and gravity, as summarized in Jurin's Law~\cite{de2013capillarity}. When a tube is inserted into passive liquids [Fig.~\ref{fig:capillary_action}a], the capillary rise is solely determined by gravity, surface tension, and the tube width.
Active liquids made of self-propelled active agents have also been shown to climb walls against gravity.
This has attracted  attention recently~\cite{wysocki2020,adkins2022dynamics,fins2024steer,mangeat2024pre,mandal2026activejurinslaw,das_capillary_2026},
but, despite these numerical and experimental efforts, a comprehensive understanding of capillary action in active fluids has
 remained out of reach, largely because surface tension in active matter is an elusive and ambiguous concept~\cite{bialke15,hermann2019,hermann2019,omar2020,lauersdorf2021phase,turci2024partial,caprini2024dynamicalclusteringwettingphenomena,solon_surprising_2025,turci2021wetting,rdc2-gsqc,besse2023prl,zhao2026wetting,thiele2026activewetting}.

Mechanical surface tension is well defined and has a clear physical interpretation. 
In scalar active matter systems with momentum-conserving interactions, a regime where mechanical pressure remains a state variable~\cite{solon_pressure15,fily17,adkins2022dynamics},  balance of forces with mechanical tension leads to an active version of the Young-Dupr\'e equation that predicts correctly the wetting angles of adsorbed droplets~\cite{zhao2026wetting}. It is thus likely that mechanical surface tension also plays a key role in active capillary action.

The debate about surface tension in active fluids was largely triggered by the fact that mechanical liquid-gas surface tension can be negative. In the context of capillary action, this would naively lead to predict a capillary drop, in contrast to the equilibrium rise. However, capillary rises, not drops, were reported in active fluids~\cite{wysocki2020,adkins2022dynamics,fins2024steer,mangeat2024pre}. Moreover, the rise depends on the tube shape [Fig.~\ref{fig:capillary_action}b,c], at odds with Jurin's law in passive systems. In particular, the scaling exponent that characterizes capillary rise appears to deviate from its passive value~\cite{wysocki2020} for an unknown reason.

In this Letter, we find that the missing key factor at the origin of capillary rise in active fluids
is the self-organized particle currents generic in scalar active matter near surfaces~\cite{nikola16,patch2018,baek18,granek2019bodies,zakine2020,dor_disordered_2022,obyrne_nonequilibrium_2023,sq25-hfsb,zhao_active_2025,metzger_equation_2026}.
Such currents have in particular been observed numerically in active wetting~\cite{mangeat2024pre,zhao2026wetting}, but their impact on capillary rise has remained unclear so far. We show that the drag they induce enters force balance and can reverse the effect of a negative mechanical surface tension. We derive an active Jurin's law from microscopics and 
show that capillary rise is not merely a local function of gravity, surface tension, and tube width, contrary to the case of passive fluids~\footnote{ The recent paper~\cite{mandal2026activejurinslaw} discusses active nematics and does not consider the effect of currents. The active nematics system does not have a well-defined mechanical surface tension, which is a different scenario from the active systems considered in this Letter.}.
Various unexpected non-local factors such as the shape of the tube and the stiffness of its outer surface significantly impact capillary rise via the modulation of circulative currents around the tube. We also find that the capillary rise
varies non-monotonically with the wetting angle. Our results demonstrate that capillary action in active systems is 
a nonlocal phenomenon that depend on the global configuration of the system, which highlights the importance and consequences of non-locality out of equilibrium.

{\it Active Brownian particles under gravity.}
\
We consider overdamped active Brownian particles moving in two dimensions, interacting via short-range pairwise repulsive forces,
subjected to a uniform gravitational force $\mathbf{g} = -g \,\mathbf{e}_y$.
We work with a large self-propulsion speed and repulsion interactions, a regime where these particles exhibit motility-induced phase separation~\cite{cates2015motility,solon2018generalized,omar2023pnas} with a negative liquid-gas surface tension $\gamma_{\rm LG}$~\cite{bialke15}. The position ${\br}_i$ and the orientation angle $\theta_i$ of the $i$-th particle evolve according to the It\^{o}-Langevin equations
\begin{align}
&\dot{\rr}_i = v_0 \mathbf{u}(\theta_i) + \mu \big( \sum_{{j\ne i}} \F_{ij} + \F_w + \mathbf{g} \big)  \label{eqn_langvin_rr} \\
&\dot{\theta}_i=\sqrt{2D_r} \, \xi_i \;, \label{eqn_langvin_theta}
\end{align}
where $v_0$ is the self-propulsion speed, $\mu=1$ is the mobility, $D_r$ is the rotational diffusion constant, and $\mathbf{u}(\theta)=(\cos\theta,\sin\theta)$ is the unit vector of the orientation angle $\theta$. The Gaussian white noise $\xi_i$ has zero mean and correlations $\langle \xi_i(t) \xi_j(t') \rangle = \delta_{ij} \delta(t - t')$. The particles interact via a harmonic repulsive potential, $\F_{ij}=-\nabla_{\rr_i}U(\rr_i-\rr_j)$ with $U(\mathbf{r})=\epsilon (r - r_0)^2 \Theta (r_0 - r)$, where $\Theta (r)$ is the Heaviside step function and $\epsilon$ is the repulsion strength. The interaction range $r_0=1$ is the length unit in the simulations.

We vertically insert a tube of width $w$ into the liquid [Fig.~\ref{fig:capillary_action}a], and model the interactions between tube and particles using a repulsive harmonic potential $\F_w=-\nabla U_w$. The mathematical expression of $U_w(\rr)$ can be found in Appendix B of the End Matter. The tube is composed of a long rectangle of width $d$ with two semicircles at both ends. Tube walls with sharp tips correspond to $d=0$. By decreasing the stiffness $\epsilon_w$ of $U_w$, the liquid transitions from fully wetting to partially wetting the tube surface.

{\it Jurin's Law of capillary action in active matter.---}
\ 
We start from the Fokker-Planck equation of Eqs.~(\ref{eqn_langvin_rr}-\ref{eqn_langvin_theta}). Integrating out the angular dependence in the time-evolution of probability density function of a single particle, detailed in~\cite{supp}, leads to the force balance equation in steady state
\begin{equation}
\nabla\cdot \bm{\sigma} - \rho\left( g \,\mathbf{e}_y + \nabla U_w\right) - \mu^{-1} \J = 0\;.\label{eqn_force_balance}
\end{equation}
The active stress tensor field $\bm{\sigma}(\rr)$ is defined by the positions and orientations of the particles, with the expression shown in Appendix A. We simulate the system in a weak gravity regime such that the stress remains isotropic away from interfaces~\footnote{As $g$ increases, the stress anisotropy in the bulk scales perturbatively as $(g/v_0)^2$~\cite{ginot2015nonequilibrium,ginot2018sedimentation}.}. The emergent particle current field is given by $\J(\rr):=\left\langle \sum_i\dot{\rr}_{i}\delta(\rr-\rr_{i})\right\rangle$.

Next, we integrate the $y$ component of Eq.~\eqref{eqn_force_balance} along the axis of the tube (dashed line in Fig.~\ref{fig:capillary_action}) between two arbitrary heights $y_1$ and $y_2$, which gives
\begin{align}
&\sigma_{yy}(0,y_2)-\sigma_{yy}(0,y_1) + \int_{y_1}^{y_2} \partial_x \sigma_{yx}(0, y)\, dy  \notag \\ 
=&g\int_{y_1}^{y_2} \rho(0, y)\, dy + \mu^{-1} \int_{y_1}^{y_2} J_y(0, y)\, dy\;.\label{eqn_integrate}
\end{align}
If the bulk of the liquid and gas phases are isotropic, $-\sigma_{yy}$ can be identified as the hydrostatic pressure.

\begin{figure}
  \begin{tikzpicture}
    \node at (-3.5,2.3) {(a)};
    \node at (0.8,2.3) {(b)};
    \node at (-2.2,0.5) {\includegraphics{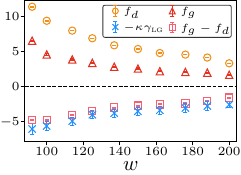}};
    \node at (2.2,0.5) {\includegraphics{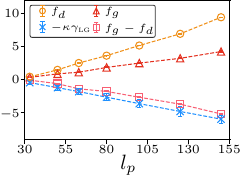}};
  \end{tikzpicture}
    \caption{Numerical verification of active Jurin's law~\eqref{eqn_active_jurin} with varying \textbf{(a)} tube width $w$ and \textbf{(b)} particles' persistence length $\ell_p=v_0/D_r$. We vary $D_r$ with fixed $v_0$ in panel (b). The contribution of gravity
    $f_g$ is defined in Eq.~\eqref{eqn_fg_def}, and the drag of currents $f_d$ is measured using $f^*_d=-\mu^{-1} \int_{-\infty}^{\infty} [J_y(0, y)-J_y(L_x/2, y)]\, dy$ where a finite size correction is considered. The difference $f_g-f_d$ is verified to be balanced by the force from the liquid-gas surface, given by $-\kappa\gamma_{\rm LG}$ measured using Eq.~\eqref{eqn_kappa_gamma}. We fix $D_r=0.05$ in panel (a) and $w=150$ in panel (b). Other parameters: $v_0=5$, $\epsilon=50$, $\epsilon_w=10$, $L_x=L_y=1024$, $g=0.015$, $d=52$.}
    \label{fig:capillary_law}
\end{figure}

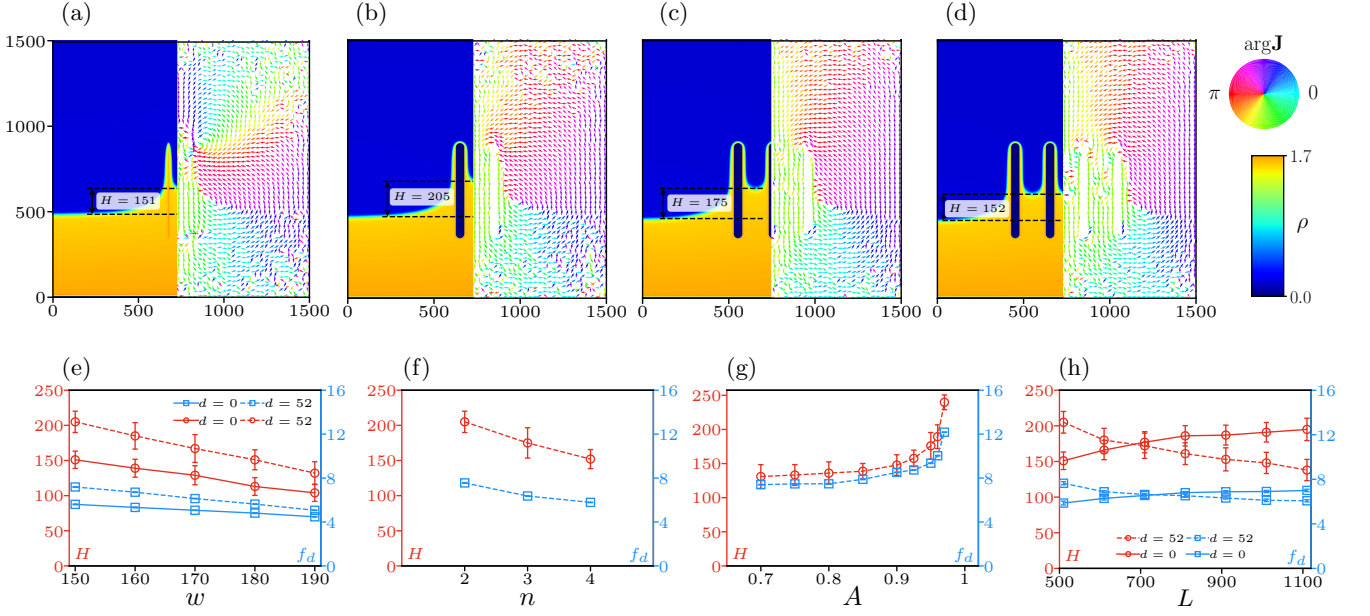
\begin{figure*}
    \centering
    \begin{tikzpicture}
        \node at (-8.4,2.2) {(a)};
        \node at (-4.5,2.2) {(b)};
        \node at (-0.5,2.2) {(c)};
        \node at (3.3,2.2) {(d)};
        \node at (-8.4,-2.5) {(e)};
        \node at (-3.9,-2.5) {(f)};
        \node at (0.4,-2.5) {(g)};
        \node at (4.8,-2.5) {(h)};
        \node at (-4.6-2.6,0.04) {\includegraphics[width=0.24\textwidth]{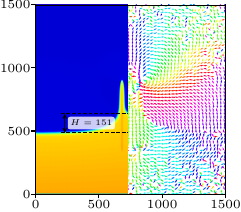}};
        \node at (-3,0) {\includegraphics[width=0.21\textwidth]{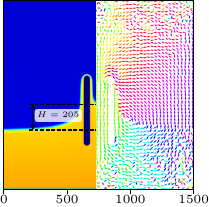}};
        \node at (0.9,0) {\includegraphics[width=0.21\textwidth]{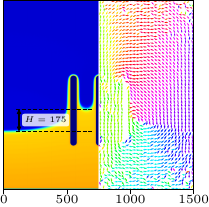}};
        \node at (4.6+0.2,0) {\includegraphics[width=0.21\textwidth]{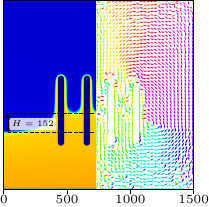}};
        \node at (7.54,-0.63)
        {\includegraphics[width=0.045\textwidth]{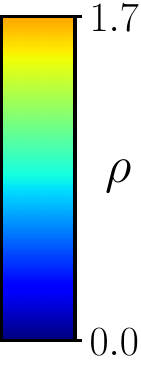}};
        \node at (7.3,1.3)
        {\includegraphics[width=0.09\textwidth]{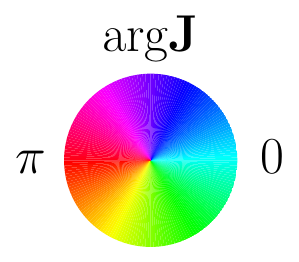}};
        \node at (-6.9,-4.2) {\includegraphics[width=0.23\textwidth]{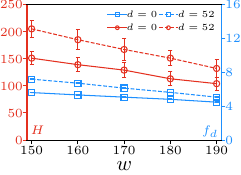}};
         \node at (-2.5,-4.2) {\includegraphics[width=0.23\textwidth]{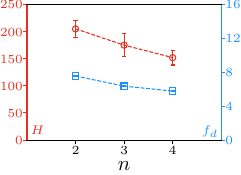}};
         \node at (1.8,-4.2) {\includegraphics[width=0.23\textwidth]{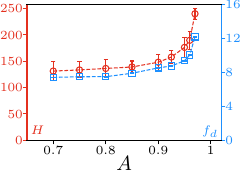}};
         \node at (6.2,-4.2) {\includegraphics[width=0.23\textwidth]{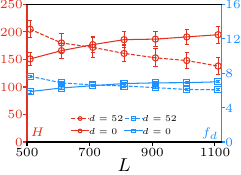}};
    \end{tikzpicture}
    \caption{Capillary action depends on the global configuration of tubes. \textbf{(a-d)} Density (left) and current (right) fields of systems with \textbf{(a)} a sharp tube, \textbf{(b)} a round tube, and \textbf{(c-d)} multiple tubes. 
    Data averaged using the left-right symmetry. Comparing (b) with (a), round tips mitigate the flow field caused by the geometric discontinuity, thereby reducing the capillary rise. In (b-d), the capillary rise decreases as the number of tubes increases. The amplitudes of the arrows are proportional to log$\vert \mathbf{J}/10^{-6}\vert$. Their color encodes their directions. 
    \textbf{(e)} Capillary rise $H$ and current drag $f_d$ with sharp tubes ($d=0$) and round tubes ($d=52$). As the tube width $w$ varies, $H$ is systematically lower in round tubes due to the decrease in $f_d$.
     \textbf{(f)} $H$ and  $f_d$ decrease as the number of tubes $n$ increases ($w=150$). 
    \textbf{(g)} $H$ and $f_d$ increase with decreasing minimal stiffness of the outer wall $\epsilon_w(1-A)/(1+A)$, where $A\to 1$ for soft walls ($w=120$, $g=0.02$, $L_x=L_y=1024$). 
    \textbf{(h)} Variation of $H$ with tube length $L$ ($w=150$, $L_x=1500$, $L_y=1500+(L-550)$). 
    In all panels, if not specified, we use $L_x=L_y=1500$, $v_0=5$, $D_r=0.05$, $g=0.015$, $d=50$, $\epsilon=50$, $\epsilon_w=10$.}
    \label{fig:tube_shape}
\end{figure*}

For a system with infinite size but finite tube height, $\sigma_{yy}(x,y)$ converges to a homogeneous profile $\tilde{\sigma}_{yy}(y)$ in the bulk of the gas ($y=y_{b,G}$) and liquid ($y=y_{b,L}$), where the translational symmetry along $x$ is recovered. We denote $\tilde{\sigma}_{yy}(y)$ as the stress measured at a position $x$ far from the tube, throughout a horizontal liquid-gas interface, so that $\tilde{\sigma}_{yy}(y):=\sigma_{yy}(x_b,y)$, where $x_b\gg w$ is a position in the bulk. Far from the tube, because the current field excited by the tube is a decaying field, the system is shear-free $\sigma_{yx}=0$ and current-free $J_y=0$. Integrating the $y$ component of Eq.~\eqref{eqn_force_balance} at $x=x_b$ then gives
\begin{equation}
    \tilde{\sigma}_{yy}(y_2)-\tilde{\sigma}_{yy}(y_1)=g\int_{y_1}^{y_2}\tilde{\rho}(y)\,dy\;, \label{eqn_profile_ansatz}
\end{equation}
where we denote $\tilde{\rho}(y)$ as the density profile at $x=x_b$. Next, we let $y_1=y_{b,L}$ and $y_2=y_{b,G}$, and replace $\sigma_{yy}(0,y_{b,L/G})=\tilde{\sigma}_{yy}(y_{b,L/G})$ in Eq.~\eqref{eqn_integrate} and~\eqref{eqn_profile_ansatz} to eliminate the stresses. Substituting Eq.~\eqref{eqn_profile_ansatz} into Eq.~\eqref{eqn_integrate} yields
\begin{equation}
    \int_{y_{b,L}}^{y_{b,G}} \partial_x \sigma_{yx}(0, y)\, dy=f_g-f_d\;,\label{eqn_active_jurin_origin}
\end{equation}
in which we denote the weight of the capillary rise,
\begin{equation}
    f_g:=g\int_{y_{b,L}}^{y_{b,G}} [\rho(0, y)-\tilde{\rho}(y)]\, dy\;, \label{eqn_fg_def}
\end{equation}
and the drag from the currents $f_d$,
\begin{equation}
    f_d:=-\mu^{-1} \int_{y_{b,L}}^{y_{b,G}} J_y(0, y)\, dy\;. \label{eqn_fd_def}
\end{equation}

In a wide tube where currents do not induce shear along $x=0$, $\partial_x \sigma_{yx}(0, y)$ is non-zero only across an interface. In~\cite{zhao2026wetting} and Appendix~A, we show that 
\begin{equation}
\int_{y_1}^{y_2} \partial_x \sigma_{yx}(0, y)\, dy=-\kappa\gamma_{\rm LG} \label{eqn_kappa_gamma}
\end{equation}
if the curvature $|\kappa|$ of the liquid-gas interface is small~\footnote{The curvature $\kappa$ is signed. We denote $\kappa<0$ for when the liquid-gas interface curves towards the liquid phase, as in Fig.~\ref{fig:capillary_action}.}, and $\gLG$ is the liquid-gas surface tension defined in Eq.~\eqref{eqn_def_gLG} on a flat interface.

Substituting Eq.~\eqref{eqn_kappa_gamma} into Eq.~\eqref{eqn_active_jurin_origin}, we obtain 
\begin{equation}
    -(d_s-1)\kappa\gamma_{\rm LG}=f_g-f_d\;,\label{eqn_active_jurin}
\end{equation}
for the spatial dimension $d_s=2$. Repeating the calculations in higher dimensions leads to Eq.~\eqref{eqn_active_jurin} for any $d_s>2$, where $\kappa$ is the mean curvature.

If gravity is weak such that the densities of the liquid and gas are nearly independent of height on the tube length scale, $f_g=(\rho_L-\rho_G)gH$ where $\rho_{L,G}$ are the densities of the liquid and gas, respectively, and $H$ is the capillary rise. Then $H$ can be solved as
\begin{equation}
    H=H_\gamma+\frac{f_d}{(\rho_L-\rho_G)g}=H_\gamma+H_J\;,\label{eqn_active_rise}
\end{equation}
where $H_J=f_d/[g(\rho_L-\rho_G)]$ is the contribution from particle currents, and the contribution from surface tension $H_\gamma$ reads
\begin{equation}
    H_{\gamma}=\frac{-(d_s-1)\kappa\gamma_{\rm LG}}{(\rho_L-\rho_G)g}\;.\label{eqn_passive_rise}
\end{equation}
In passive systems, $H_J=0$, and Eq.~\eqref{eqn_active_rise} reduces to the classic Jurin's law. The new term $H_J$ induces violations of Jurin's law and brings new physics to the capillary action of active matter. Equation~\eqref{eqn_active_rise} is the central result of this Letter.

We verify the construction of Eq.~\eqref{eqn_active_jurin} in particle simulations [Fig.~\ref{fig:capillary_law}]. We numerically solve the dynamics~\eqref{eqn_langvin_rr} and~\eqref{eqn_langvin_theta} in systems with periodic boundaries in the $x$ direction and confined in the $y$ direction by horizontal walls [Fig.~\ref{fig:capillary_action}]. For simplicity, we consider hard tubes where the liquid fully wets the walls. Eq.~\eqref{eqn_active_jurin} is verified while changing the tube width and the persistence length [Fig.~\ref{fig:capillary_law}]. In all cases, we notice that the contribution from the surface tension is negative, while $H>0$ is due to a large $f_d>0$ that counters the effects of the surface tension~\footnote{In simulations our system has a finite size, and the currents thus do not fully vanish at the boundaries. A finite size correction is applied to the numerical measurements on $f_d$,
$
    f^*_d=-\mu^{-1} \int_{-\infty}^{\infty} [J_y(0, y)-J_y(L_x/2, y)]\, dy\;.
$
The difference between the blue and pink symbols in Fig.~\ref{fig:capillary_law} is accounted for by the non-vanishing currents at the top and bottom of the system. We show in~\cite{supp} that, while perfect agreement can be achieved by taking these finite size effects into account, they decrease as the system size increases.}.

The active Jurin's law Eq.~\eqref{eqn_active_rise} suggests that $H\sim g^{-1}$ in the limit of weak gravity $g\to 0$, since $\gLG$, $f_d$ and $\rho_L-\rho_G$ should converge to a finite value in the $g\to 0$ limit. In simulations, we measured an exponent close to $-0.77$~\cite{supp}, accompanied by a significant finite-size contribution that decays with $g$. The reported exponent~\cite{wysocki2020} that differs from $-1$ is likely due to similar finite-size effects.

{\it Global currents modify capillary rise.---}
\ 
We now demonstrate the counterintuitive consequences of the drag of $f_d$ in the active Jurin's law.
For passive systems in weak gravity, the capillary rise is given by Eq~\eqref{eqn_passive_rise}, where the curvature $\kappa$ is determined by the width of the tube $w$ and the Young's wetting angle $\varphi$, which are properties of the inner wall of the tube. Since all quantities in Eq.~\eqref{eqn_passive_rise} are located inside the tube, the global shape of the tube has no effect on the capillary rise $H_\gamma$ in passive systems. 

However, the active capillary rise is governed by Eq.~\eqref{eqn_active_jurin}. Given the negative surface tension $\gamma_{\rm LG}<0$, $H_\gamma<0$ leads to a capillary drop, which means that $H_J$ is solely responsible for the observed capillary rise. Physically, this suggests that the capillary rise emerges from circulating currents due to particles constantly pumping into the tube walls. The particle currents depend on the detailed structure of the tube, including its outer wall. This suggests that $H$ is no longer a local quantity within the tube but is affected by the global shape of the tube in active systems.

To demonstrate the impact of the global shape of the tube, we first compare the capillary rise of active fluids in tubes with sharp tips and with round tips in simulations [Fig.~\ref{fig:tube_shape}a,b]. The two tubes have the same wall stiffness and width, which would result in the same capillary rise 
$H_\gamma$ in a passive fluid. However, we observe a significant increase in $H$ in the tube with round tips. By measuring the particle currents, we find that the sharp tip, as also considered in~\cite{wysocki2020}, introduces a singular point that exerts a large force on the active liquid and excites a complex vortex near the tip. The vortex current acts as a ratchet pump that simultaneously injects particles into the tube and creates flows of particles above the tube. With the specific tube length in Fig.~\ref{fig:tube_shape}a,b, the net drag of the complex currents created by the sharp tips is lower than that created by the round tips, and the corresponding capillary rise systematically reduces as the width of the tube $w$ varies [Fig.~\ref{fig:tube_shape}e]. Later we will show that both $H$ and $f_d$ scale differently in round and sharp tubes when the length of the tube varies.

To directly demonstrate the impact of pumping particles, we progressively add more tubes with the same stiffness to form a comb-like object 
[Fig.~\ref{fig:tube_shape}b-d,f]. In passive systems, the capillary rises, set by Eq.~\eqref{eqn_passive_rise}, would be identical in all tubes. In the active case, the main contribution to capillary rise comes from the currents created outside the comb. Distributing these injected particles across multiple tubes leads to a decrease in $H$ as the number of tubes increases [Fig.~\ref{fig:tube_shape}b-d]. We note that $H$ is not expected to be inversely proportional to the number of tubes, because the tips of the inner tube walls and the climbing liquids inside the tubes also contribute to particle currents, which are independent of the number of tubes.

The tips of the tube are not the only sources of circulating currents that account for the capillary rise. Active fluids climb the outer wall of the tube and form a capillary bridge, which can create vortex currents under gravity. To demonstrate its impact, we change the stiffness of the outer wall and find that the capillary height varies [Fig.~\ref{fig:tube_shape}g]. To avoid discontinuous changes in wall stiffness, which can create more complex vortex currents (see \cite{supp}), we consider a smooth potential for the outer wall that interpolates between the stiffness of the tip $\epsilon_w$ and the stiffness of the outer wall $\epsilon_w(1-A)/(1+A)$, where $A$ controls the minimal stiffness. With $A$ increasing to 1, the outer wall becomes softer, and the wetting angle $\varphi$ increases away from 0 (fully wetting). A finite wetting angle then induces additional currents~\cite{zhao2026wetting}, which forces more particles into the tube and increases $H$ [Fig.~\ref{fig:tube_shape}g]. 

Finally, we find that $H$ varies with tube length $L$. In passive systems, $H$ is independent of $L$ when $L>H$. For active fluids in round tubes, $H$ continuously decreases even beyond $L>2H$ [Fig.~\ref{fig:tube_shape}h], while the opposite is observed in sharp tubes. To understand this phenomenon, we first note that the currents induced by external forces are dipolar and generally decay as $r^{-2}$, with $r$ being the distance from its source~\cite{baek18,granek2019bodies,granek2023inclusions}. Then, we decompose the contribution to $H$ of different current sources, $H_J=H_{\rm tip}+H_{\rm out}$, where $H_{\rm tip}$ denotes the contribution from the currents excited by the tips, and $H_{\rm out}$ is from other sources outside the tube. The flux of particles induced by the tips is independent of $L$, while that induced by other sources scales as $L^{-2}$. Since the drag in the tube scales as $\bar{J}L$, where $\bar{J}$ is the mean flux of particles inside the tube, $H_{\rm tip}$ scales as $L$ while $H_{\rm out}$ scales as $L^{-1}$. This results in a setup-dependent scaling of $H$ with $L$. For the round tubes, the currents created by the tips are weak. Thus, $H$ is dominated by $H_{\rm out}$ and decreases with $L$. With $L\to\infty$, it is likely that $H$ becomes negative due to $H_\gamma$, even though this regime is beyond our numerical capability. For the sharp tips, $H_{\rm tip}$ has a significant contribution. If the drag of currents above the tube is kept constant (see the discussions on Fig.~\ref{fig:tube_shape}a and e), $H$ increases with $L$.~\footnote{To ensure that the flux in the tube and the drag of currents above the tube remain constant in finite-size systems, we fix the distance between the tube tip and the upper boundary of the system. Thus, $L_y$ increases as $L$ increases, and $L_y-L$ is kept constant in Fig.~\ref{fig:tube_shape}h. The contribution in $f_d$ from the currents outside the tube is kept constant, so that $H_{\rm tip}$ increases with $L$.} Notably, the difference in $H$ of the sharp and round tubes persists as $L\to\infty$.

{\it Conclusion.---}
\
We investigated the capillary action of active fluids in tubes under uniform and weak gravity. By explicitly constructing an active Jurin's law, we identify the self-organized ratchet currents forcing particles into the tube as
the new ingredient responsible for the capillary rise in an active fluid with a negative mechanical liquid-gas surface tension, a phenomenon at odds with expectations drawn from passive systems. These currents depend on the global configuration of the system. As a result, capillary rise is no longer a local quantity determined solely by tube width, surface tensions, and gravity. We showed that the shape and number of the tubes, the stiffness of their outer walls, and other unexpected quantities have a significant impact on the capillary rise. 

Our theory applies to dry active systems with momentum-preserving interactions. So far we presented cases with fluids fully wetting the tube. In Appendix~C we show that our theory also applies when fluids partially wet the tube, and the currents induced by the contact angle leads to more complex phenomena. Systems of active particles coupled to momentum-conserving fluids will be an interesting direction for future research. Although our theory will have to be complemented by new source and sink terms in systems where pressure and surface tension do not obey an equation of state, our qualitative conclusions are likely to apply to these systems as well: we expect the capillary rise to be significantly influenced by particle currents in all active fluids where they arise.

Finally, spontaneous currents are a defining feature of non-equilibrium systems in general, and scalar active matter in particular, that is attracting growing attention~\cite{granek2023inclusions,li_robust_2024,langford_phase_2025}. They depend on the global configuration of the system and lead to non-local couplings in the hydrodynamic modes. They can either break or create expectations of universality that we sometimes implicitly build from equilibrium intuition. 
Our findings suggest that a full account of wetting phenomena in most, if not all, active systems should include non-local effects in some form, e.g., by explicating the dynamics of the orientation field or by including convolution terms.

\begin{acknowledgments}
We thank Hugues Chat\'e, Xiaqing Shi, Alexandre Solon, and
Julien Tailleur for inspiring discussions. YZ acknowledges support from National Natural Science Foundation of China (Grant 12304252).
\end{acknowledgments}

\bibliography{biblio_surface}
\onecolumngrid

\vspace{12pt}
\noindent\hrulefill \hspace{24pt} {\bf End Matter} \hspace{24pt} \hrulefill
\vspace{12pt}

\twocolumngrid
{\it Appendix A: Definitions of the stress tensor and liquid-gas surface tension.---}
The stress tensor in the force-balance equation~\eqref{eqn_force_balance} is given by~\cite{zhao2026wetting}
\begin{align}
    \sigma_{\alpha\beta}(\rr)=&\sigma_{\alpha\beta}^{\rm IK}(\rr)-\mu^{-1}\frac{v_0}{D_r}\left\langle\sum_i\dot{\rr}_i\otimes\uu_i\delta(\rr-\rr_i)\right\rangle , \label{eqn_def_sigma}\\
    \sigma_{\alpha\beta}^{\rm IK}(\rr)=&-\frac{1}{2}\left\langle \sum_{i,j}\F_{ij}\otimes\rr_{ij}\int_0^1ds\,\delta(\rr-s\rr_{ij}-\rr_j)\right\rangle ,
\end{align}
where $\sigma_{\alpha\beta}^{\rm IK}$ is the standard Irving-Kirkwood stress tensor induced by the pairwise interactions between the particles. $\rr_{ij}=\rr_{i}-\rr_{j}$, and $\F_{ij}=-\nabla_{\rr_i}U(\rr_{ij})$ is the force from the particle $j$ to $i$. Tensor product is denoted as $\otimes$.

\begin{figure}
  \begin{tikzpicture}
    \node at (-5-1,2.5) {(a)};
    \node at (0-1.6,2.5) {(b)};
    \node at (-2.2-2.5,0.65) {\includegraphics[width=0.25\textwidth]{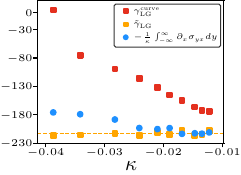}};
    \node at (2.2-2.7,0.58) {\includegraphics[width=0.25\textwidth]{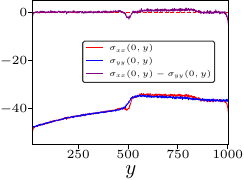}};
  \end{tikzpicture}
    \caption{(a) $\gamma_{\rm LG}^{\rm curve}$ measured inside the tube on curved interface with respect to the curvature $\kappa$. It is compared with $\tilde{\gamma}_{\rm LG}$ measured at the boundary of the system, where the curvature is negligible. The dashed line shows the average of $\tilde{\gamma}_{\rm LG}$. The blue dots denote the phenomenological surface tension $\gLG^*$. (b) The distribution of $\sigma_{xx}$ and $\sigma_{yy}$ along the central axis of the tube, which exhibits strong anisotropy under the influence of the flow field, the red dashed line indicates $x=0$, the tube width $w=100$. Parameters: $v_0=5$, $D_r=0.05$, $L_x=L_y=1024$, $g=0.015$.}
    \label{fig:kappa_gamma}
\end{figure}

The liquid-gas surface tension $\gamma_{\rm LG}$ is defined by the anisotropic part of the stress tensor across a flat interface~\cite{zhao2026wetting}
\begin{equation}
\gamma_{\rm LG}=\int_{h_0-l}^{h_0+l}dy^{\prime}[\sigma_{xx}(0,y^{\prime})-\sigma_{yy}(0,y^{\prime})]\;,\label{eqn_def_gLG}
\end{equation}
where we assume the liquid-gas interface is parallel to the $x$ axis and located at $h_0$, and $l$ is a length longer than the thickness of the liquid-gas interface. We note that Eq.~\eqref{eqn_def_gLG} applies when the curvature of the interface is negligible. In Fig.~\ref{fig:kappa_gamma}a, we show that the measurements of Eq.~\eqref{eqn_def_gLG} at the system boundary $x=\pm L_x/2$, denoted as $\tilde{\gamma}_{\rm LG}$, are stable across systems. However, $\gLG^{\rm curve}$ measured inside the tube on a curved interface shows a rather large deviation. While Eq.~\eqref{eqn_kappa_gamma} holds for small to medium $\kappa$ [Fig.~\ref{fig:kappa_gamma}a], applying Eq.~\eqref{eqn_def_gLG} on the strongly curved interface inside the tube leads to a value $\gLG^{\rm curve}$ with a large deviation from $\gLG$, which fails to satisfy Eq.~\eqref{eqn_kappa_gamma}. For large $|\kappa|$, the strong currents create anisotropy in the gas inside the tube [Fig.~\ref{fig:kappa_gamma}b], and Eq.~\eqref{eqn_kappa_gamma} does not hold. But it still demonstrates a contribution of a curved interface, where we nevertheless denote a phenomenological surface tension $\gamma^*_{\rm LG}=-\kappa^{-1}\int_{y_1}^{y_2} \partial_x \sigma_{yx}(0, y)\, dy$, and the physics of capillary action in active systems is not qualitatively altered. Thus, in Fig.~\ref{fig:capillary_law}, we measure $-\kappa\gLG$ using Eq.~\eqref{eqn_kappa_gamma}.

{\it Appendix B: Definitions of wall potentials.---}
The potential of a tube wall with round tips takes the form 
\begin{equation}
    U_w(\rr)=U_{w,\mathrm{L}}(\rr)+U_{w,\mathrm{R}}(\rr)+U_{w,\rm T}(\rr)+U_{w,\rm B}(\rr)\;.
\end{equation}
We denote $x_{w,\rm L}$ and $x_{w,\rm R}$ as the left and right positions of the tube wall, respectively, and $y_{c,\rm B}$ and $y_{c,\rm T}$ as the vertical positions of the centers of the semicircular tips. For round tips, we require $x_{w,\rm R}-x_{w,\rm L}=d$ where $d$ is the diameter of the semi-circles. For $y_{w,\rm B}<y<y_{w,\rm T}$, $U_{w,\mathrm{L}}(\rr)$ and $U_{w,\mathrm{R}}(\rr)$ are non-zero, and
\begin{align}
    U_{w,\mathrm{L}}(\rr)=&\epsilon_{w}(x-x_{w,L})^2\Theta(x-x_{w,L})\;, \\
    U_{w,\mathrm{R}}(\rr)=&\epsilon_{w}(x-x_{w,R})^2\Theta(x_{w,R}-x)\;.
\end{align}
For $y<y_{c,\rm B}$ or $y>y_{c,\rm T}$, $U_{w,\mathrm{T}}$ and $U_{w,\mathrm{B}}$ are nonzero if $|\rr-\rr_{w,{\rm T/B}}|<d/2$,
\begin{equation}
    U_{w,\mathrm{T/B}}(\rr)=\epsilon_{w}\left(|\rr-\rr_{w,\rm {T/B}}|-\tfrac{d}{2}\right)^2\;,
\end{equation}
and $\rr_{w,\rm T/B}=(x_{w,{\rm T/B}},y_{c,\rm T/B})$ are the centers of the semi-circles. For sharp tips, we simply take $U_{w,\mathrm{T}}=U_{w,\mathrm{B}}=0$, and $x_{w,\rm R}-x_{w,\rm L}=2$. Note that the object is always repulsive in active systems. Then, a tube is composed of two walls separated by a distance $w$ [Fig.~\ref{fig:capillary_action}].

In Fig.~\ref{fig:tube_shape}g, the potential of the outer walls are replaced by
\begin{equation}
    U_{w,\mathrm{outer}}(x,y)=\frac{\epsilon_{w}[1+A\cos(2\pi \tilde{y}/L)]}{1+A}(x-x_w)^2\;,
\end{equation}
where $L=y_{c,\rm T}-y_{c,\rm B}$ is the tube length [Fig.~\ref{fig:capillary_action}], and $\tilde{y}=y-y_{c,\rm B}$. 

\begin{figure}
  \begin{tikzpicture}
      \node at (-6.5,2.5) {(a)};
      \node at (-2,2.5) {(b)};
        \node at (-2.2-3.2,0.5) {\includegraphics[width=0.21\textwidth]{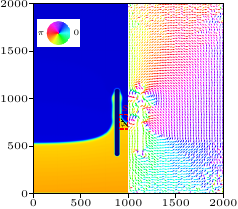}};
        \node at (-3.1,0.56)
        {\includegraphics[width=0.047\textwidth]{jet_colorbar_thin.pdf}};
        \node at (2.2-2.6,0.5) {\includegraphics[width=0.25\textwidth]{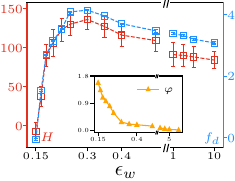}};
    \end{tikzpicture}
   \caption{Capillary rise in active systems depends non-monotonically on the contact angle. (a) Density (left) and current (right) fields of capillary action in partially-wetting case. (b) The capillary rise $H$ and the drag of currents $f_d$ as a function of the wall stiffness $\epsilon_w$ of the tube. Inset shows the contact angle $\varphi$ as a function of $\epsilon_w$. Parameters: $v_0=5$, $D_r=0.05$, $g=0.015$, $w=200$, $\epsilon=50$. In panel (a), $L_x=L_y=2000$. In panel (b), $L_x=L_y=1024$.}
    \label{fig:partial_wetting_tubes}
\end{figure}

{\it Appendix C: Capillary action with  partially wetting liquids.---} 
When the liquid only partially wets the tube walls, the wetting angle $\varphi$ departs from zero [Fig.~\ref{fig:partial_wetting_tubes}a]. 
In passive systems, this leads to $\kappa=-2\cos\varphi/w$, and the capillary rise is
\begin{equation}
    H_\gamma^p=\frac{2(d_s-1)\gLG\cos\varphi}{(\rho_L-\rho_G)wg}\;,\label{eqn_capillary_rise_passive_partial}
\end{equation}
which decreases when $\varphi$ increases from 0.
In active systems, partial wetting occurs when the inner tube wall is soft enough. The negative sign of $\gLG$ then suggests that $H$ increases when the liquid partially wets the tube. This is indeed the case for small $\varphi$ [Fig.~\ref{fig:partial_wetting_tubes}b]. However, a non-zero wetting angle in active systems is always accompanied by the presence of vortex currents~\cite{zhao2026wetting}, which contribute to capillary rise. Thus, the relationship between $H$ and $\varphi$ in active systems is more complex. As $\varphi$ increases further, $H$ decreases again and changes sign, as it does in passive fluids with positive $\gLG$ [Fig.~\ref{fig:partial_wetting_tubes}b].

To be specific, the active Jurin's law~\eqref{eqn_active_jurin} remains valid, but the vortex currents near the contact angle deform the circular shape of the meniscus and deviate $\kappa$ from $-2\cos\varphi/w$. To relate $\kappa$ with $w$, we integrate the $y$ component of the force-balance equation~\eqref{eqn_force_balance} in the region $\mathcal{S}=[-w/2,0]\times[y_1,y_2]$ [Fig.~\ref{fig:partial_wetting_tubes}a], which reads
\begin{align}
    &\gLG\cos\varphi+\int_{-\frac{w}{2}}^{0}dx\,\left[\sigma_{yy}(x,y_2)-\sigma_{yy}(x,y_1)\right] \non
    &-g\int_{\mathcal{S}} dxdy\,\rho=\mu^{-1}\int_{\mathcal{S}} dxdy\,J_y\;,\label{eqn_int_force_in_tube}
\end{align}
where we have used the facts that the interface is parallel to the $x$ axis at $x=0$, $\sigma_{yx}(0,y)=0$, and~\cite{zhao2026wetting}
\begin{equation}
    -\int_{y_1}^{y_2}dy\,\sigma_{yx}\left(-\frac{w}{2},y\right)=\gLG\cos\varphi\;.
\end{equation}

Next, integrating the $y$ component of Eq.~\eqref{eqn_force_balance} along $x=0$ from $y_1$ to $y_2$ yields
\begin{align}
    &-\kappa\gLG+\sigma_{yy}(0,y_2)-\sigma_{yy}(0,y_1)-g\int_{y_1}^{y_2} dy\,\rho(0,y)\non
    &=\mu^{-1}\int_{y_1}^{y_2} dy\,J_y(0,y)\;.\label{eqn_int_force_in_axis}
\end{align}
If the tube length $L$ is long enough, we can choose $y_1$ and $y_2$ to be inside the tube with negligible currents, and
\begin{equation}
    \int_{-\frac{w}{2}}^{0}dx\,\sigma_{yy}(x,y_{1,2})=\frac{w}{2}\sigma_{yy}(0,y_{1,2})\;.\label{eqn_int_sigmayy}
\end{equation}
Multiplying Eq.~\eqref{eqn_int_force_in_axis} by $-w/2$ and adding with Eq.~\eqref{eqn_int_force_in_tube}, we have
\begin{align}
    &\gLG\cos\varphi+\tfrac{1}{2} \kappa \gLG w + F_d^{\rm AYD} \non
    &-g\int_{\mathcal{S}} dxdy\,[\rho(x,y)-\rho(0,y)]=0\;, \label{eqn_contact_angle_and_kappa}
\end{align}
where
\begin{equation}
    F_d^{\rm AYD}=-\mu^{-1}\int_{\mathcal{S}}dxdy\,\left[J_y(x,y)-J_y(0,y)\right]\label{eqn_Fd_AYD}
\end{equation}
is the drag of currents near the contact angle with respect to a background flow $J_y(0,y)$, which is related to partial wetting but is not the force entering the Young-Dupr\'e equation~\cite{zhao2026wetting}. The gravitation term in Eq.~\eqref{eqn_contact_angle_and_kappa} corresponds to the liquid in the curved part of the meniscus. If the volume of liquid curved by wetting the tube is much smaller than $\gLG/(g\rho_L)$, this term can be neglected. Finally, 
using Eq.~\eqref{eqn_contact_angle_and_kappa}, we obtain
\begin{equation}
    \kappa=-\frac{2\cos\varphi}{w}-\frac{2}{w\gLG}F_d^{\rm AYD}\;. \label{eqn_correct_kappa}
\end{equation}
The $F_d^{\rm AYD}$ contribution leads to a new term in $H$:
\begin{equation}
    H=H_\gamma^p+H_J+\frac{2F_d^{\rm AYD}}{(\rho_L-\rho_G)wg}=H_\gamma^p+H_J+H_J^{\rm AYD}\;,\label{eqn_active_rise_partial_wetting}
\end{equation}
where $H_\gamma^p$ is the passive contribution with partial wetting, given by Eq.~\eqref{eqn_capillary_rise_passive_partial}. The contribution from the contact angle in the tube is denoted as $H_J^{\rm AYD}=2F_d^{\rm AYD}/[(\rho_L-\rho_G)wg]$.

We note that Eq.~\eqref{eqn_active_rise_partial_wetting} is valid in the limit of infinite system size and tube width with weak gravity and is more prone to finite size effects than in fully-wetting systems. We note that a finite thickness $\ell_b$ of the fluid-solid boundary on the tube wall has to be considered, and $w$ is replaced by an effective tube width $w-2\ell_b$. (See~\cite{supp} for more discussions on  finite-size effects.) We expect, as $\varphi$ increases, $H_{\gamma}^p$ to increase monotonically, and $H_J^{\rm AYD}$ to decrease monotonically from 0. As in the fully wetting case, $H_J$ depends on the contact angle and on the condensation droplets on the \textit{outer} wall, leading to complex behavior.


\end{document}